\pdfoutput=1 

\documentclass[aps,preprint,showpacs,preprintnumbers,amsmath,amssymb]{revtex4}

\usepackage{graphicx}
\usepackage{dcolumn}
\usepackage{bm}

\begin{document}


\title{Fingerprinting Soft Materials: A Framework for Characterizing \\ Nonlinear Viscoelasticity}

\author{Randy H. Ewoldt}
\author{Gareth H. McKinley}
\author{A. E. Hosoi}
\email{peko@mit.edu}
\affiliation{%
Hatsopoulos Microfluids Laboratory, Department of Mechanical Engineering, Massachusetts Institute of Technology, Cambridge, Massachusetts 02139, USA
}%

\date{October 26, 2007}

\begin{abstract}
We introduce a comprehensive scheme to physically quantify both viscous and elastic rheological nonlinearities simultaneously, using an imposed large amplitude oscillatory shear (LAOS) strain.  The new framework naturally lends a physical interpretation to commonly reported Fourier coefficients of the nonlinear stress response.  Additionally, we address the ambiguities inherent in the standard definitions of viscoelastic moduli when extended into the nonlinear regime, and define new measures which reveal behavior that is obscured by conventional techniques.
\end{abstract}

\maketitle

Biopolymer networks \cite{Gardel04,Storm05,Fletcher07}, wormlike micelles \cite{Spenley93}, colloidal gels \cite{Gisler99}, and metastable soft solids in general \cite{Wyss07}, can be classified as   nonlinear viscoelastic materials and as such have been of interest to experimentalists for many decades (e.g.~\cite{Philippoff66}). The biological and industrial processes associated with these materials often involve large deformations, yet standard methods of characterizing their nonlinear rheological properties rely on techniques designed for small strains.  In this Letter, we develop a new and systematic framework for quantifying the {\em nonlinear} viscoelastic response of soft materials which enables us to describe a unique ``rheological fingerprint'' of an \emph{a priori} unknown substance.

Both the elastic and viscous characteristics of a material can be examined simultaneously by imposing an oscillatory shear strain, $\gamma(t)=\gamma_0 \sin(\omega t)$, which consequently imposes a phase-shifted strain-rate $\dot\gamma(t)$.  Here $\omega$ is the imposed oscillation frequency, $\gamma_0$ is the maximum strain amplitude and $t$ is time.  At small strain amplitudes when the response is linear, the material is commonly characterized by the viscoelastic moduli $G'(\omega)$, $G''(\omega)$, as determined from the components of the stress in phase with $\gamma(t)$ and $\dot\gamma(t)$, respectively.  For a purely elastic linear solid, the elastic modulus $G'$ is equivalent to the shear modulus $G$.
Similarly, for a purely viscous Newtonian fluid with viscosity $\mu$, the loss modulus $G''=\mu\omega$.  However, these viscoelastic moduli are not uniquely defined once the material response becomes nonlinear, since higher order harmonics emerge.
\begin{figure}
\includegraphics[width=5.5in]
{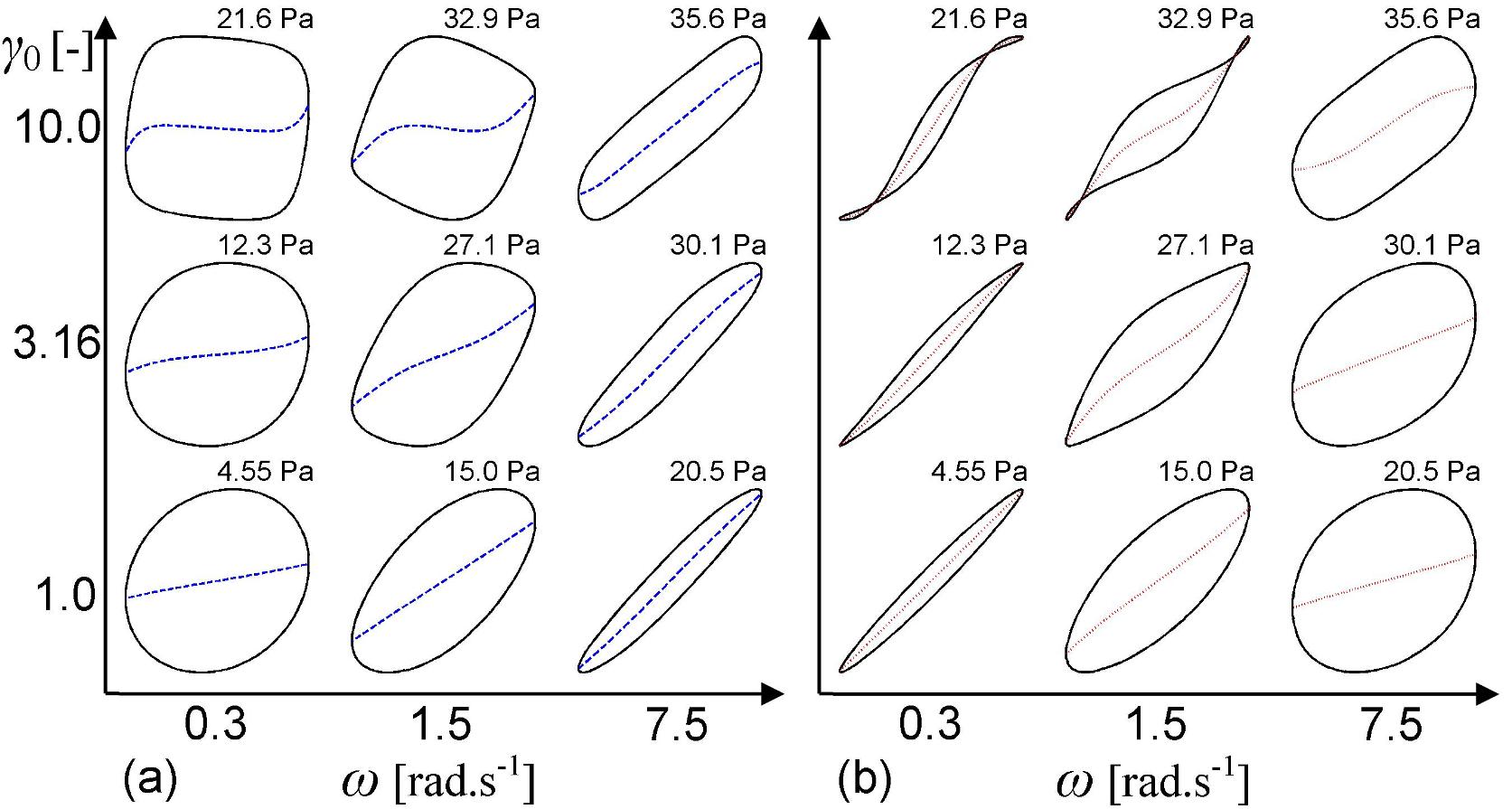}
\caption{Viscoelastic material response from nine experimental oscillatory tests of the micellar solution. Each trajectory is positioned according to the imposed values $\{\omega,\gamma_0\}$.  (a) Elastic Lissajous curves: solid lines are total stress $\sigma(t)/\sigma_{max}$ vs.~$\gamma(t)/\gamma_0$, dashed lines are elastic stress $\sigma'(t)/\sigma_{max}$ vs.~$\gamma(t)/\gamma_0$; (b) Viscous Lissajous curves: solid lines are total stress $\sigma(t)/\sigma_{max}$ vs.~$\dot\gamma(t)/\dot\gamma_0$, dotted lines are viscous stress $\sigma''(t)/\sigma_{max}$ vs.~$\dot\gamma(t)/\dot\gamma_0$ . The maximum stress, $\sigma_{max}$, is indicated above each curve.}
\label{fig:CPyCl-Pipkin}
\end{figure}
For convenience the moduli are often determined by the coefficients of the first harmonic, $G_1'$ and $G_1''$ (see Eqn.~\ref{eq:FTwhole:sub1}).  These measures of the viscoelastic moduli are arbitrary and often fail to capture the rich nonlinearities that appear in the raw data signal \cite{Ewoldt07SM}.

An example of such rich behavior is shown in the large amplitude oscillatory shear (LAOS) results from a wormlike micelle solution in Fig.~\ref{fig:CPyCl-Pipkin}.  The periodic stress response $\sigma(t;\omega,\gamma_0)$ at steady state is plotted against either $\gamma(t)$ or $\dot\gamma(t)$, the simultaneous phase-shifted inputs.  These parametric plots are commonly called Lissajous curves (or more accurately, Bowditch-Lissajous curves \footnote{Orbital trajectories bearing J. A. Lissajous' name were studied by N. Bowditch in 1815, predating Lissajous' treatment in 1857 as summarized by \cite{Crowell81}}).  In this parameter space, a linear viscoelastic response appears as an ellipse which is progressively distorted by material nonlinearity.  We refer to the $\sigma(t)$ vs.~$\gamma(t)$ curves (Fig.~\ref{fig:CPyCl-Pipkin}a) as \emph{elastic} Lissajous curves to distinguish them from the \emph{viscous} Lissajous curves (Fig.~\ref{fig:CPyCl-Pipkin}b) which plot $\sigma(t)$ as a function of the shear-rate $\dot\gamma(t)$.

The most common method of quantifying LAOS tests is Fourier transform (FT) rheology \cite{Wilhelm02}.  For a sinusoidal strain input $\gamma(t)=\gamma_0 \sin \omega t$, the stress response can be represented as a Fourier series
\begin{eqnarray}\label{eq:FTwhole}
\sigma \left( {t;\omega ,\gamma _0 } \right) = \gamma _0 \sum\limits_{n\text{ odd}}  \{ &G_n'\left( {\omega ,\gamma _0 } \right)\sin n\omega t&   \nonumber\\
+ &G_n''\left( {\omega ,\gamma _0 } \right)\cos n\omega t&\} .
\label{eq:FTwhole:sub1}
\end{eqnarray}
This expression emphasizes elasticity.  Viscous scaling results by factoring out $\dot\gamma_0=\gamma_0\omega$ rather than $\gamma_0$, in which case the coefficients are $\eta_n''=G_n'/\omega$ and $\eta_n'=G_n''/\omega$.
In the linear regime $\eta'$ is known as the dynamic viscosity.  Only odd-harmonics are included in this representation since the stress response is symmetric with respect to shear strain or shear-rate, i.e.~the material response is the same in both shear directions.  Even-harmonic terms can be observed in transient responses, secondary flows \cite{Atalik04}, or dynamic wall slip \cite{Graham95}, but these conditions will not be considered here.  Although this FT framework is mathematically robust and reduces to the linear viscoelastic framework in the limit of small strains, it lacks a physical interpretation of the higher-order coefficients.

Other methods have been used to quantify viscoelastic nonlinearities \cite{Tee75,Janmey83,Cho05,Klein07}, however these techniques either lack physical interpretation, cannot be calculated uniquely, or do not apply simultaneously to both elastic and viscous phenomena.  For example, a recently proposed decomposition \cite{Klein07} uses sets of sine, square, and triangular waves to describe a nonlinear response.  These basis functions are not orthogonal and thus blur measurement of the linear viscoelastic response and qualitative interpretations of the progressive onset of nonlinearity.  Similarly, the stress decomposition of \cite{Cho05} suffers from non-orthogonality, as outlined below.  It is therefore desirable to develop a complete and systematic framework for quantifying nonlinear viscoelasticity which avoids these ambiguities.

To illustrate our proposed new framework, we apply it to a wormlike micelle solution and a natural biopolymeric hydrogel, gastropod pedal mucus.  The wormlike micelle solution (prepared as in \cite{Rothstein07}) consists of cetylpyridinium chloride (CPyCl) and sodium salicylate (NaSal) dissolved in brine.  The ratio of CPyCl/NaSal is 100 mM/50 mM (3.2 wt\%/0.76 wt\%) in a 100 mM (0.56 wt\%) NaCl aqueous solution.  Pedal mucus samples were collected as in \cite{Ewoldt07SM}.  All experiments were performed on a strain-controlled ARES rheometer (TA Instruments) equipped with a Peltier plate maintained at $T=22^\text{o}$C, using a solvent trap to inhibit evaporation.  The micellar solution was tested with a cone-plate geometry (diameter $D=50$~mm, angle $\alpha=2.3^\text{o}$).  The mucus was tested with a plate-plate configuration
($D=8$~mm and gap $h=550$~$\mu$m).  To eliminate slip, the plate-plate surfaces were covered with adhesive-backed waterproof sandpaper, 600 grit (ARC Abrasives Inc.).

To interpret the data, we extend the method of orthogonal stress decomposition \cite{Cho05}, which uses symmetry arguments to decompose the generic nonlinear stress response into a superposition of an elastic stress $\sigma'(x)$, where $x=\gamma/\gamma_0=\sin\omega t$, and viscous stress $\sigma''(y)$ where $y=\dot \gamma/\dot \gamma_0=\cos\omega t$.
Thus, plotting $\sigma'$ vs.~$x$ or  $\sigma''$ vs.~$y$ produces single-valued functions in contrast to the closed loops formed by total stress $\sigma$ vs.~$\gamma$ or $\sigma$ vs.~$\dot\gamma$ (see Fig.~\ref{fig:CPyCl-Pipkin}).  The \emph{intra}-cycle elastic and viscous nonlinearities (i.e.~nonlinearities within a given steady state cycle) are therefore related to the nonlinearity of these functional forms.  Cho et al.~\cite{Cho05} suggest a polynomial regression fit to these lines of elastic and viscous stress.  However, the material properties represented by these coefficients are not unique since they depend on the number of fitting coefficients arbitrarily chosen by the user.

Instead, we suggest that these curves be represented by the set of Chebyshev polynomials of the first kind as they are symmetric, bounded and orthogonal on the finite domain, $-1\le x \le 1$, and can be easily converted to the Fourier coefficients which have dominated the discussion on quantitative LAOS analysis.
Using this basis set, the elastic and viscous contributions to the measured stress response can be written as
\begin{subequations}
\label{eq:ChebyshevDefWhole}
\begin{eqnarray}
\sigma'(x) = \gamma_0 \sum\limits_{n\text{ odd}} {e_n \left( {\omega ,\gamma_0} \right)T_n (x)}
\label{eq:ChebyshevDefWhole:sub1}
\\
\sigma''(y) = \dot\gamma_0 \sum\limits_{n\text{ odd}} {v_n \left( {\omega ,\gamma_0} \right)T_n (y)}
\label{eq:ChebyshevDefWhole:sub2}
\end{eqnarray}
\end{subequations}
where $T_n(x)$ is the $n$th-order Chebyshev polynomial of the first kind.   We refer to $e_n$  as the elastic Chebyshev coefficients and $v_n$ as the viscous Chebyshev coefficients.

In the linear regime Eq.~\ref{eq:ChebyshevDefWhole} recovers the linear viscoelastic result such that $e_1=G_1'$ and $v_1=\eta_1'=G_1''/\omega$.  We interpret any initial deviation from linearity, i.e.~the $n=3$ harmonic, as follows.  A positive contribution of the third-order polynomial $T_3(x)=4x^3-3x$ results in a higher stress at maximum strain, $x\rightarrow 1$,
than represented by the first-order contribution alone.  Thus, $e_3>0$ corresponds to \emph{intra}-cycle strain-stiffening of the elastic stress, whereas $e_3<0$ indicates strain-softening.  Similarly, a positive value for $v_3$ represents \emph{intra}-cycle shear-thickening
and $v_3<0$ describes shear-thinning.  Note that this framework gives a direct physical interpretation of the commonly reported third-order Fourier coefficients $G_3'$, $G_3''$ \cite{Wilhelm02,ReimersDealy96}.
As defined in Eq.~\ref{eq:ChebyshevDefWhole}, the Chebyshev coefficients can be related back to the Fourier coefficients of Eq.~\ref{eq:FTwhole}.  Using the identity $T_n(\cos\theta)=\cos(n\theta)$, together with $\sin\theta=\cos (\pi/2-\theta)$ to show that $T_n \left( {\sin \theta } \right) = \sin \left( {n\theta } \right)\left( { - 1} \right)^{\frac{{n - 1}}
{2}}$ (for $n$ odd), the relationship between Chebyshev coefficients in the strain or strain-rate domain and Fourier coefficients in the time domain is given by
\begin{subequations}
\label{eq:ChebyFTrelation}
\begin{eqnarray}
e_n  &=& G_n '\left( { - 1} \right)^{\frac{{n - 1}}
{2}}
\label{eq:ChebyFTrelation:sub1}
\\
v_n &=& \frac{{G_n''}} {\omega } = \eta _n '
\label{eq:ChebyFTrelation:sub1}
\end{eqnarray}
\end{subequations}
(for $n$ odd).  Thus, just as the third-order Chebyshev coefficients have the physical interpretation described above, so too the third-order Fourier coefficients (with appropriate sign corrections for $G_3'$) give physical insight into the nature of deviation from linear viscoelasticity.  
When applied to the micelle data of Fig.~\ref{fig:CPyCl-Pipkin}, the
Chebyshev coefficients offer physical insight to the \emph{intra}-cycle nonlinearities.  For $\omega=0.3$~rad.s$^{-1}$ and $\gamma_0=\{1, 3.16, 10\}$, the results are $e_3=\{0.023,0.127,0.235\}$~Pa and $v_3=\{-0.011,-0.079,-0.288\}$~Pa.s, indicating progressive \emph{intra}-cycle strain-stiffening in the elastic response ($e_3>0$) combined with \emph{intra}-cycle shear-thinning in the viscous response ($v_3<0$).

This physical interpretation of higher order coefficients motivates several new geometric measures for reporting the magnitudes of first-order (linear) viscoelastic moduli in the nonlinear regime, to complement the often reported first-order Fourier coefficients.  These additional measures all reduce to $G'$ and $G''$ in the linear regime, but diverge systematically when used to analyze a nonlinear signal, offering additional physical insight (beyond that captured by the average measures $G_1'$, $G_1''$) into the underlying rheological response.  The variation in these new measures can be reported as a function of imposed strain amplitude $\gamma_0$ to indicate the nature of the nonlinearity across different steady state cycles (\emph{inter}-cycle nonlinearities).

\begin{figure}
\includegraphics[width=2.75in]
{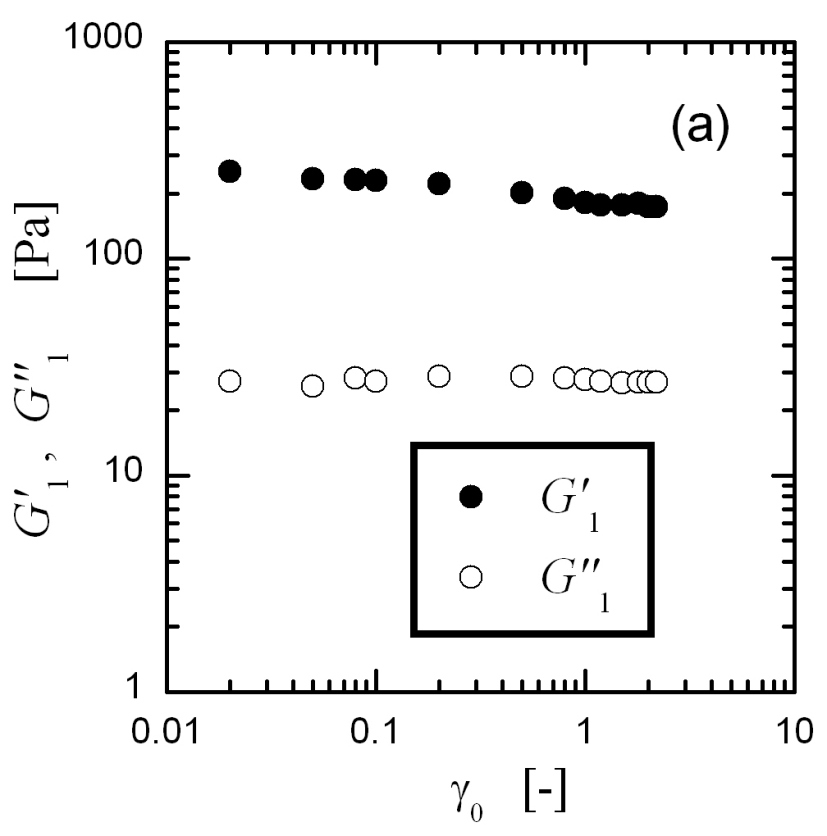}
\includegraphics[width=2.75in]
{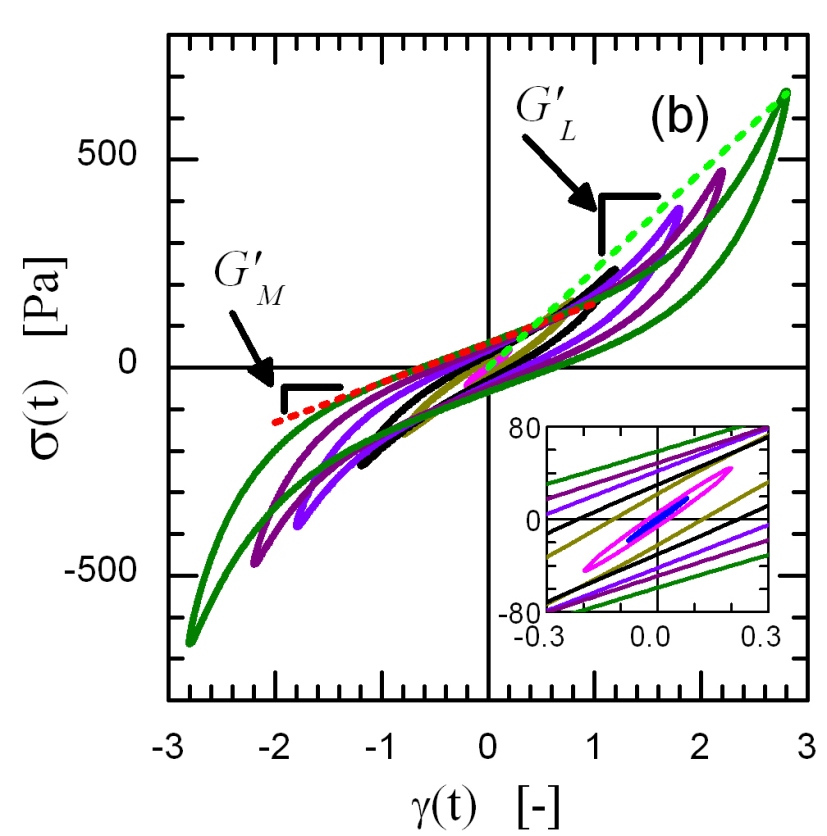}
\caption{\label{fig:SlimeGpLiss} Oscillatory strain sweeps of pedal mucus from \emph{Limax maximus} at a frequency $\omega=3$~rad.s$^{-1}$. (a) Typical rheometer output of the fluid viscoelasticity as parameterized by the first-order Fourier moduli.  (b) Plotting the raw data from every-other point as $\sigma(t)$ vs.~$\gamma(t)$, with graphical representation of elastic moduli $G_M'$,$G_L'$ for $\gamma_0=2.8$.}
\end{figure}

The nonlinear viscoelastic response of native pedal mucus gel secreted by the terrestrial slug \emph{Limax maximus} provides a striking example of the shortcomings of the conventional FT rheology framework \cite{Ewoldt07SM}.  When tested in oscillatory shear, the ``elastic modulus" reported by the instrument (i.e.~the first harmonic elastic modulus, $G_1'$) decreases slightly with strain amplitude, implying a weak strain-softening.  However, examining the raw data in the form of Lissajous curves indicates a pronounced local strain-stiffening within a given steady-state cycle at sufficiently large stress amplitudes \cite{Ewoldt07SM}.  A series of strain-controlled oscillatory tests at a fixed frequency of $\omega=3$~rad.s$^{-1}$ showing this peculiar nonlinear behavior of pedal mucus is given in Fig.~\ref{fig:SlimeGpLiss}.
The elastic Lissajous curves in Fig.~\ref{fig:SlimeGpLiss}b are elliptical for small $\gamma_0$ (see inset), but become progressively distorted in the nonlinear regime. At large strains, the shear stress is greater than one would expect by projecting the center portion of the ellipse, suggesting elastic strain-stiffening which is not captured by the first harmonic elastic modulus.  This behavior is not unique to pedal mucus but appears to be common in soft biological materials
and can been seen, for example, in data reported by \cite{MaWirtz99} for a keratin filament network.  In both cases \emph{intra}-cycle strain-stiffening is readily apparent in the elastic Lissajous curves even though the familiar ``viscoelastic moduli'' do not appear to increase as a function of strain amplitude.

Reporting an ``elastic modulus" of such a material as $G_1'$ is misleading, as other harmonics may also store energy \cite{Ganeriwala87}.  The first-harmonic represents a sine transform, $G_1'=\omega/(\pi\gamma_0^2)\oint {\sigma(t)\gamma(t)dt}$, which is a measure of average elasticity, and is unable to distinctly represent the local elastic response of a material at small and large strains.  To capture this local behavior, we define a set of geometrically-motivated elastic moduli and derive their relation to the Fourier and Chebyshev coefficients. Consider the following:
\begin{eqnarray}
G_M ' \equiv \left. {\frac{{d\sigma }}
{{d\gamma }}} \right|_{\gamma  = 0} = \sum\limits_{n\text{ odd}} {nG_n '} = e_1 - 3e_3 + ...\label{eq:GMdef}
\\
G_L ' \equiv \left. {\frac{\sigma }
{\gamma }} \right|_{\gamma  = \gamma _0 } = \sum\limits_{n\text{ }\,\text{odd}} {G_n '\left( { - 1} \right)^{\frac{{n - 1}}
{2}} } = e_1 + e_3 + ...\label{eq:GLdef}
\end{eqnarray}
where $G_M'$ is the \emph{minimum-strain modulus} and $G_L'$ is the \emph{large-strain modulus}.  These measures can be visualized graphically as shown in Fig.~\ref{fig:SlimeGpLiss}b by the broken lines.  The measures are deliberately chosen such that both converge to the linear elastic modulus $G'$ in the limit of small strains.
We also define a set of dynamic viscosities for reporting the viscous or dissipative response.  The definitions and relation to the Fourier series of Eq.~\ref{eq:FTwhole} are
\begin{eqnarray}
\eta_M' \equiv \left. {\frac{{d\sigma }} {{d\dot \gamma }}} \right|_{\dot \gamma  = 0} = \sum\limits_{n\text{ odd}} {n \frac{G_n''}
{\omega }\left( { - 1} \right)^{\frac{{n - 1}} {2}} } = v_1 - 3v_3 + ...
\label{eq:EtaMdef}
\\
\eta_L' \equiv \left. {\frac{\sigma } {{\dot \gamma }}} \right|_{\dot \gamma  = \dot \gamma _0 }
= \sum\limits_{n\text{ odd}} {\frac{G_n''}{\omega }} = v_1 + v_3 + ...
\label{eq:EtaLdef}
\end{eqnarray}

\begin{figure}
\includegraphics[width=2.75in]
{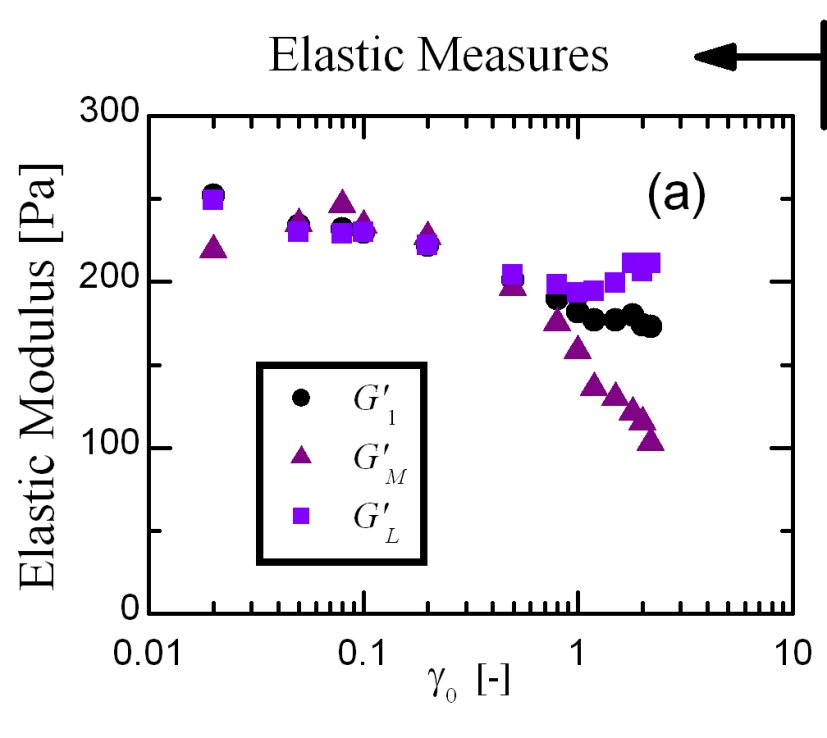}
\includegraphics[width=2.75in]
{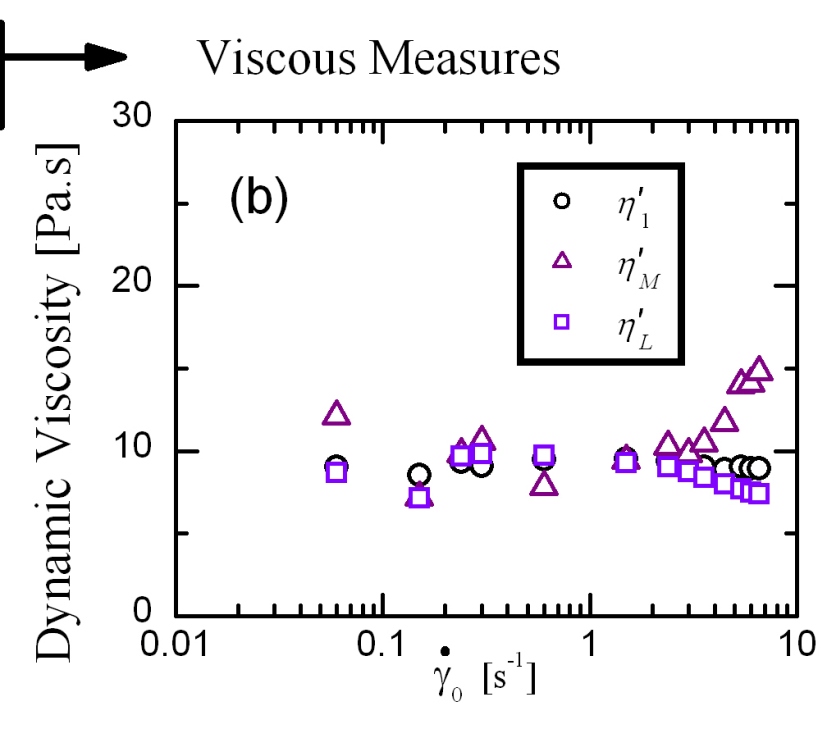}\\
\includegraphics[width=2.75in]
{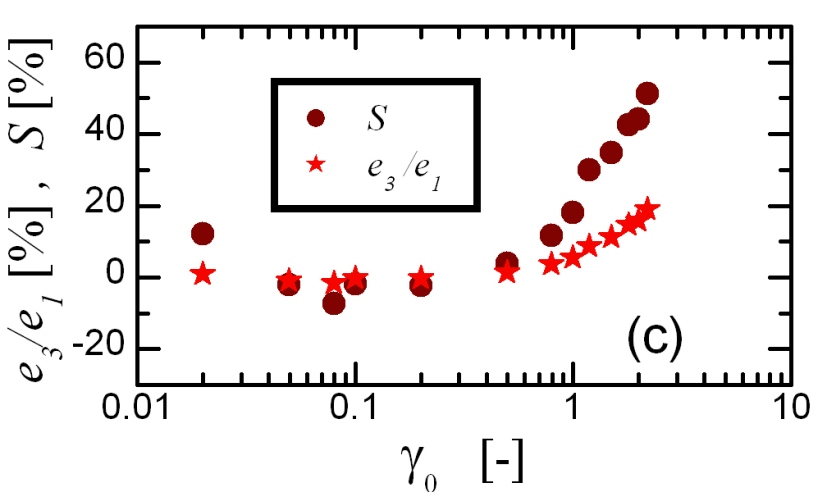}
\includegraphics[width=2.75in]
{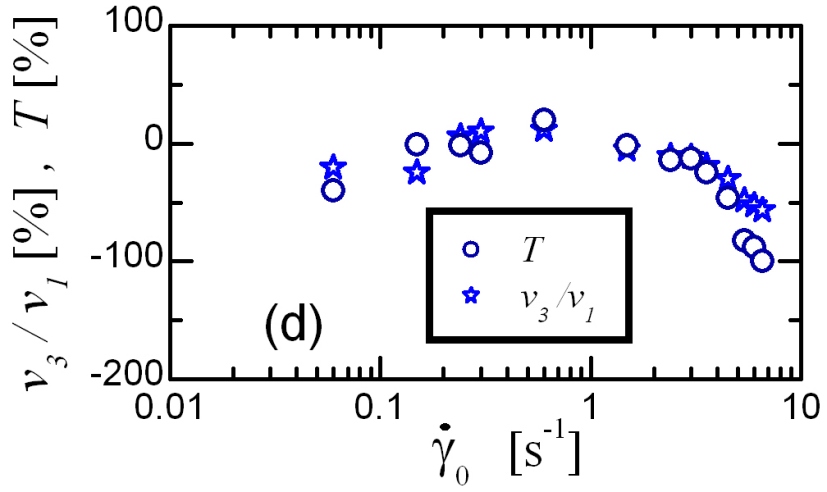}
\caption{\label{fig:SlimeMITlaos} Oscillatory shear test from Fig.~\ref{fig:SlimeGpLiss}, analyzed within the new framework. (a) Elastic moduli: minimum-strain and large-strain elastic moduli compared to first harmonic elastic modulus. (b) Dynamic viscosities: minimum-rate and large-rate dynamic viscosities compared to first harmonic dynamic viscosity. (c) Elastic nonlinearity measures: scaled 3rd order elastic Chebyshev coefficient $e_3/e_1$ and strain-stiffening ratio $S$, both indicate \emph{intra}-cycle strain-stiffening   (d) Viscous nonlinearity measures: scaled 3rd order viscous Chebyshev coefficient $v_3/v_1$ and shear-thickening ratio $T$, both indicate \emph{intra}-cycle shear-thinning.}
\end{figure}

The \emph{inter}-cycle variations of these new measures, as applied to pedal mucus, are reported in Fig.~\ref{fig:SlimeMITlaos}.  Here the minimum-strain modulus softens, $G_M'$ decreasing with $\gamma_0$, whereas the large-strain elasticity first softens then stiffens, $G_L'$ first decreasing then increasing with $\gamma_0$ (see Fig.~\ref{fig:SlimeMITlaos}a).  From Fig.~\ref{fig:SlimeMITlaos}b it is also apparent that the dissipation at small strain-rates ($\eta_M'$) increases with increasing $\dot \gamma_0$, whereas the dissipative nature at large strain-rates (represented by $\eta_L'$) decreases from one cycle to the next.  This rich behavior is obscured by the average measures of elastic modulus $G_1'$ and dynamic viscosity $\eta_1'=G_1''/\omega$ which are commonly reported.

The \emph{intra}-cycle nonlinearity which distorts the linear viscoelastic ellipse can also be quantified by comparing these new material measures.  Here we define the strain-stiffening ratio as
\begin{equation}
S \equiv \frac{G_L'-G_M'}{G_L'} = \frac{4e_3 + ...}{e_1 + e_3 + ...}
\end{equation}
where $S=0$ for a linear elastic response, $S>0$ indicates \emph{intra}-cycle strain-stiffening, and $S<0$ corresponds to \emph{intra}-cycle strain-softening.  Users of this framework may also choose to compare the moduli using e.g.~the ratio $G_L'/G_M'=(1-S)^{-1}$; we chose the former for convenience in relating the new measures to higher order Fourier/Chebyshev coefficients, and to eliminate potential singularities as $G_M'\rightarrow0$.  We similarly define the shear-thickening ratio as
\begin{equation}
T \equiv \frac{\eta_L'-\eta_M'}{\eta_L'} = \frac{4v_3 + ...}{v_1 + v_3 + ...}
\end{equation}
where $T=0$ indicates a single harmonic linear viscous response, $T>0$ represents \emph{intra}-cycle shear-thickening, and $T<0$ \emph{intra}-cycle shear-thinning.
The \emph{intra}-cycle nonlinearities of pedal mucus are quantified in Fig.~\ref{fig:SlimeMITlaos}c,d.  Both $S$ and $e_3$ are positive at the largest strain amplitudes, indicating strain-stiffening, while the nonlinear viscous measures $T$ and $v_3$ are both negative at large strain-amplitudes, indicating \emph{intra}-cycle shear-thinning.

In conclusion, the growing interest in biological and other soft materials compels a need for consistent, quantitative, low-dimensional descriptions of nonlinear material responses. We propose a new comprehensive framework to quantify such nonlinear viscoelastic behavior. The scheme provides a physical interpretation of deviations from linear viscoelastic behavior, describing elastic and viscous nonlinearities separately, simultaneously, and more thoroughly than currently reported measures (summarized by example in Fig.~\ref{fig:SlimeMITlaos}).  In addition, the method can be easily applied to previously collected data using existing software \footnote{Software to analyze raw measurement of \{$\gamma(t),\sigma(t)$\} is available from the authors or from http://web.mit.edu/nnf/}.  These new measures may lend insight in the development of constitutive models, and may also serve as a more sensitive test for comparing the distinguishing features of different materials.  The framework is broadly applicable to any complex fluid or soft material which can be tested in oscillatory shear, and serves as a complement to the familiar and successful linear viscoelastic framework embodied in $G'(\omega)$ and $G''(\omega)$.


\bibliography{PRLrefs}

\end{document}